\documentclass[twocolumn,showpacs,preprintnumbers,amsmath,amssymb,aps]{revtex4}

\usepackage{graphicx}
\usepackage{amssymb}
\usepackage{epstopdf}
\usepackage{bm}% bold math

\begin{document}
    
\title{Hamilton-Jacobi Many-Worlds Theory and the Heisenberg Uncertainty Principle}
    
\author{F.~J.~Tipler}
\affiliation{Department of Mathematics and Department of Physics, Tulane University, New Orleans, LA 70118}

\date{\today}

\begin{abstract}
I show that the classical Hamilton-Jacobi (H-J) equation can be used as a technique to study quantum mechanical problems.  I first show that the the Schr\"odinger equation is just the classical H-J equation, constrained by a condition that forces the solutions of the H-J equation to be everywhere $C^2$.  That is, quantum mechanics is just classical mechanics constrained to ensure that ``God does not play dice with the universe.''  I show that this condition, which imposes global determinism, strongly suggests that $\psi^*\psi$ measures the density of universes in a multiverse.    I show that this interpretation implies the Born Interpretation, and that the function space for $\psi$ is larger than a Hilbert space, with plane waves automatically included.   Finally, I use H-J theory to derive the momentum-position uncertainty relation, thus proving that in quantum mechanics, uncertainty arises from the interference of the other universes of the multiverse, not from some intrinsic indeterminism in nature.

\end{abstract}

\maketitle

\section{Quantum Mechanics as Deterministic Classical Mechanics}

Consider the action for $N$ particles, with $3N$ configuration space coordinates $x_i$:

\begin{equation}
S(x_i,t) = \int_{t_0}^t L(x_i, \dot x_i, t)\,dt
 \label{eq:action}
\end{equation}	
 
 \noindent
Let $S(x_i,t)$ be a function taken along configuration space paths starting from any fixed position at time $t_0$, but with the final position and final time $t$ to be variables.  
 
Taking the first variation of the action integral (\ref{eq:action}) gives
 
 \begin{equation}
\delta S = \sum_i\left(\left[\frac{\partial L}{\partial {\dot x_i}}\delta x_i\right]^t_{t_0}   +   \int_{t_0}^t \left(\frac{\partial L}{\partial x_i} - \frac{d}{dt} \frac{\partial L}{\partial {\dot x_i}}\right)\delta x_i\,dt\right)
 \label{eq:actionfirstvariation}
\end{equation}	
 
If we require that the paths we consider obey the laws of physics, then the expression inside the integral that multiplies the variation $\delta x_i$ must vanish for each $i$ (Euler-Lagrange equations), giving

 \begin{equation}
\delta S = \sum_i\frac{\partial L}{\partial {\dot x_i}}\delta x_i(t) =  \sum_i p_i\delta x_i(t)
 \label{eq:firstvariationendpoint}
\end{equation}
 
\noindent
where I have used the assumption, that, although the particles could have started from any point $x_i$ at the initial time, the variation of that initial point is zero: $\delta x_i(t_0) = 0$.  I have also used the fact that $\partial L/\partial {\dot x_i}$ is the momentum $p_i$ conjugate to the coordinate $x_i$.
 
 But considering the action to be a function of the final positions $x_i(t)$ and final time $t$, we also have for the first variation
 
 \begin{equation}
\delta S = \sum_i\frac{\partial S}{\partial {x_i}}\delta x_i(t)
 \label{eq:otherfirstvariationendpoint}
\end{equation}
 
 Comparing the two expressions (\ref{eq:firstvariationendpoint}) and (\ref{eq:otherfirstvariationendpoint}) for the first variation of the action we have
 
\begin{equation}
p_i = \frac{\partial S}{\partial x_i}
 \label{eq:momentumS}
\end{equation}

Differentiating the action with respect to time, we obtain
 
 \begin{equation}
\frac{dS}{dt} = L = \frac{\partial S}{\partial t} + \sum_i\frac{\partial S}{\partial x_i}{\dot x_i}
 \label{eq:actionderivative}
\end{equation}	

Substituting (\ref{eq:momentumS}) into equation (\ref{eq:actionderivative}) we obtain

 \begin{equation}
L = \frac{\partial S}{\partial t} + \sum_ip_i{\dot x_i}
 \label{eq:actionderivative2}
\end{equation}

\noindent
which can be written

 \begin{equation} \frac{\partial S}{\partial t} + \left[\sum_ip_i{\dot x_i} - L\right] = 0
 \label{eq:actionderivative3}
\end{equation}
 
 The expression in brackets is recognized as the Hamiltonian $H(x_i, p_j, t)$, so using (\ref{eq:momentumS}) we can write equation (\ref{eq:actionderivative3}) as
 
  \begin{equation}
  \frac{\partial S}{\partial t} + H\left(x_i, \frac{\partial S}{\partial x_i}, t\right) = 0
 \label{eq:HJequation}
\end{equation}
 
 \noindent
which is an equation for Hamilton's principle function $S$ which we see is also the action.  Equation (\ref{eq:HJequation}) is course the Hamilton-Jacobi equation.  This derivation of the H-J equation, from the principle of least action in its most general sense, is adapted from a related derivation in \cite{LandauLifshitzNewton}.

One can also derive the H-J equation by using a canonical transformation \cite{Goldstein} to give a Hamiltonian $K$ with $K=0$  In this alternative derivation, $S$ is the generating function of a canonical transformation, and the generating function $S(x_i, P_i, t)$ is assumed to be a function of the original coordinates $x_i$, and new momentum $P_j$, which, since $K$ is zero, are constants of the motion.  In this case, the original momenta $p_i$ satisfy $p_i =\partial S/\partial x_i$.

From either derivation, it follows that the trajectory of a particle is that set of trajectories which are everywhere normal to the surface $S= {\rm constant}$.  However, the function $S$ actually gives an infinity of trajectories, rather than a single trajectory.  Consider the simplest case, a single particle with vanishing potential.  Then a solution is $S = p_0x - Et$, which corresponds to a particle moving in the $+x$ direction, with constant momentum $p_0$ and constant energy $E$.  There are an infinity of such particle trajectories, one for each point in the $yz$-plane.  In this case, the trajectories never interact, so each trajectory behaves as if the others did not exist.  

Let us consider the Hamiltonian 

  \begin{equation}
H =  \sum_{i=1}^k\frac{({\vec\nabla_i}  S)^2}{2m_i}+ V(x_1, x_2, \ldots, x_{3N}, t)
 \label{eq:generalHamiltonian}
\end{equation}

\noindent
where there are $k$ particle types, each type with mass $m_i$, and each particle type has $l_i$ particles.  Thus $\sum_{1=1}^k l_i = N$, where $N$ is the total number of particles.

There are potentials $V(x_1, x_2, \ldots, x_{3N}, t)$ in which the trajectories intersect.  Consider a spherically symmetric attractive $1/r$ potential, and consider a wave that far from the center of attraction is the plane wave $S = p_ox - Et$.  The trajectories will bend toward each other, and intersect on the opposite side of the center of the potential.  If we were an observer traveling on any of the initial plane wave trajectories, the collision of the other trajectories would establish the real existence of the other trajectories. 

This way of establishing the existence of an entity has been called the ``Dr. Johnson theory of existence \cite{Boswell}'' --- something is said to ``exist'' if it can hit us  --- but it is merely both the common sense theory of existence, and the standard way to prove the existence of a particle in physics.  If we run into a wall in the darkness, the existence of the wall is established.  If a neutrino is observed to hit an atom, the existence of neutrinos is established.  No longer do we doubt the existence of neutrinos even though most neutrinos from the Sun pass through the Earth without interacting.  

So we are forced to assume the existence of all the trajectories of the plane wave, even if the actual potential is such that they never intersect our own trajectory.

For any particular solution of the H-J equation, an individual trajectory I shall call a {\it universe} and the collection of all trajectories I shall term the {\it multiverse}.  The space of all globally $C^2$ solutions to the H-J equation is the space of all {\it possible} multiverses.  If we have the particular solution to the H-J equation that yields the trajectory of our actual universe, then by the Dr. Johnson theory of existence, the other trajectories in the thus defined multiverse do also, and so we refer to this particular {\it possible} multiverse, as the ``actual multiverse,''

For arbitrary potentials, the intersection is usually too ``hard."  If the trajectories intersect, then the assumption that the action function $S(x_i, t)$  is everywhere differentiable is violated.  Recall that the actual particle trajectories are those which are everywhere normal to the constant $S$ surfaces.  If two trajectories intersect, then at the point of intersection, there are two normals, hence the normal at that point is not defined.  This means that Hamilton's principle function $S$  --- the action --- is no longer $C^1$, since $\vec\nabla_i S$ is not defined at the point of intersection.   Points of the wave $S$ where $\vec\nabla_i S$ is not defined is called a ``caustic'' of the wave.

This violation of differentiability can be completely avoided by adding to every potential an additional potential $U(x_i, t)$, with

 \begin{equation}
U =  -\left(\frac{\hbar^2}{
2}\right)\sum_{i=1}^k \frac{1}{m_i}\left(\frac{\nabla_i^2{\rm R}}{{\rm R}}\right)
 \label{eq:quanpot}
\end{equation}	
 
\noindent
where as before there are $k$ particle types, each type with mass $m_i$, and each particle type has $l_i$ particles.  Thus as before $\sum_{1=1}^k l_i = N$, where $N$ is the total number of particles.  The Laplace operator $\nabla^2_i$ is, for each $i$, the Laplace operator in $3l_i$ dimensions.  The symbol $\hbar$ is some constant with the dimension of action.  The function $\rm R$ is a function of all the coordinates of configuration space, and the time $t$, and thus with $N$ particles,  it will be a function of $3N + 1$ variables.  The function ${\rm R}$ satisfies the continuity equation

 \begin{equation}
\frac{\partial{\rm R}^2}{\partial t} + \sum_{i=1}^k\vec\nabla_i\cdot
\left({\rm R}^2\frac{{\vec\nabla_i}  S}{m_i}\right)=0
\label{eq:Bohm1}
\end{equation}	

I have called (\ref{eq:Bohm1}) a ``continuity equattion''  since $\frac{{\vec\nabla_i}  S}{m_i}$ is the generalized velocity.  We now have two equations, the H-J equation with potential $V + U$, and equation (\ref{eq:Bohm1}).  Notice that since the original potential was arbitrary except depending only on the coordinates of configuration space, and not on the time derivatives of these occridinates $\dot x_i$, adding the new potential $U$ subject to (\ref{eq:Bohm1}) does not change the fact that the new H-J equation satisfies the generalized least action principle, and follows from a canonical transformation.

These two equations, the H-J equation with potential $V + U$, and equation (\ref{eq:Bohm1}), can be combined into a single equation if we define a function $\psi$ by the expression

\begin{equation}
\psi \equiv {\rm R}\exp(i{ S}/\hbar)
\label{eq:Pilotwave}
\end{equation}	

Then the function $\psi$ is easily seen to satisfy the single equation for the complex valued function $\psi$:

\begin{equation}
i\hbar\frac{\partial\psi}{ \partial t} = -\frac{\hbar^2}{
2}\left[ \sum_{i=1}^k\frac{\nabla_i^2\psi}{m_i}\right] + V(x_1, x_2, \ldots, x_{3N}, t)\psi
 \label{eq:QM}
\end{equation}
 
Since equation (\ref{eq:QM}) is linear, it cannot give rise to caustics, and hence is globally $C^2$.  Since it is equivalent to the pair of classical equations, they also are globally $C^2$.  Since the potential (\ref{eq:quanpot}) is added to the usual H-J equation to ensure the global solutions are globally $C^2$, we can regard the potential (\ref{eq:quanpot}) as a ``smoothing potential.''

The addition of the smoothing potential to the H-J equation is analogous to Maxwell's addition of the displacement current to Ampere's Law.  Both are added to enforce mathematical consistency of the underlying theory.  Both additions involve a continuity equation, the charge conservation equation in Maxwell's case, and equation (\ref{eq:Bohm1}) in the Hamilton-Jacobi case.  Both additions have profound experimental consequences.  In particular, the mathematical consistency of Hamilton-Jacobi mechanics implies that there must exist a fundamental constant $\hbar$ with the units of action.
 
Suppose we were to solve equation (\ref{eq:QM}) for a single electron and a single proton (thus $k=2$, $l_i =1$ for both $i$, and $3N = 6$) and  assume these two particles were bound by the usual attractive electrostatic potential.  If we realized that the difference in energies between two particular solutions with different energies were equal to the energies of the Balmer series photons, we would equate the constant of action $\hbar$ to the reduced Planck's constant.

With this value of the unknown constant of action $\hbar$, equation (\ref{eq:QM}) is just the general Schr\"odinger equation for $N$ spinless particles with potential $V$.  Bohm \cite{Bohm52a}, \cite{Bohm52b} and Landau (\cite{Landau77}, p. 51--52) starting with Sch\"odinger's equation for a single particle of mass $m$, obtained the classical H-J equation potential  $V +U$ and the continuity equation (\ref{eq:Bohm1}).  Here I have started with the completely general classical H-J theory, and derived the generalized Schr\"odinger equation by enforcing the assumption that $S(x_i,t)$ be globally $C^2$ by adding the potential (\ref{eq:quanpot}).  I conjecture that in some sense, the {\it ensatz} I have used here is a unique way of enforcing global differentiability.  (it may be that a different  {\it ensatz} would correspond to a different operator factor ordering, a problem that does not arise for non-relativistic quantum mechanics with potential.)

Only if the action function $S$ is globally differentiable is the H-J equation globally deterministic.  So we can regard quantum mechanics as resulting from requiring that classical mechanics be globally deterministic.  Quantum mechanics thus arises from insisting that ``God does not play dice with the universe.''

Now we must determine what the function $\rm R$ represents.  We introduced the function $\rm R$ in order to prevent trajectory intersections from forming singularities, which is to say, from preventing the trajectory number density from becoming infinite.  This suggests that $\rm R$ has something to do with defining a number density of universes in the multiverse.  Equation (\ref{eq:Bohm1}) is exactly the continuity equation we would expect if $\rm R^2$ was proportional to the number density of universes in the multiverse.  Thus let us assume $\rm R^2$ is  indeed proportional to the number of universes in the multiverse.  Notice that the number density can be defined only up to a constant, if the Schr\"odinger equation is to remain linear.  

Consider first a single particle.   If $\nabla^2{\rm R} = 0$, then the potential $U$ vanishes, and we are left with only the usual H-J equation.  Such a solution is appropriate when we have trajectories which never intersect, which means that we can completely ignore the existence of the other trajectories beside our own.  We can act as if the other universes of the multiverse do not exist.  We can call these trajectories ``classical trajectories,'' although ''non-interacting'' would be better, since {\it all} trajectories are classical: we have never left classical mechanics.  

Indeed, the space of all multiverses is larger than the space of ``quantum mechanics,''  if we try to restrict the latter space merely to all functions in a Hilbert space.   A $\psi$ with $\nabla^2{\rm R} = 0$ is allowed in classical mechanics, but it is not an element in a Hilbert space. 

For by Liouville's theorem, the only non-singular solutions to $\nabla^2{\rm R} = 0$ are ${\rm R = constant}$, and $\rm R$ for which $\lim_{|x| \rightarrow +\infty}{\rm |R|} = \infty$.  Neither class of solutions would define a $\psi$ in a Hilbert space.  

For solutions with  ${\rm R = constant}$, the continuity equation becomes

\begin{equation}
\sum_{i=1}^k\frac{1}{m_i}\nabla_i^2 S = 0
\label{eq:Rconstantcontunuity}
\end{equation}	

If the variables separate, then for each $m_i$, we would have

\begin{equation}
\nabla_i^2 S = 0
\label{eq:LaplaceS}
\end{equation}	

\noindent
which tells us, once again by Liouville's Theorem, that either $S = {\rm constant}$, or $\lim_{|x| \rightarrow +\infty}{ |S|} = \infty$.  The plane wave $S = p_0x - Et$ is an example of the latter type of solution.

In the language of $\psi = {\rm R}e^{S/\hbar}$, which is to say in the language of Schr\"odinger's equation, the plane wave $S = p_0x - Et$ is represented as

\begin{equation}
\psi = {\rm R}_0\exp(i[p_0x - Et]/\hbar)
\label{eq:planewave}
\end{equation}	

\noindent
where $ {\rm R}_0$ is a constant, expressing explicitly that the density of trajectories of universes in the multiverse is a constant.

In Schr\"odinger language, we see that one can obtain the momentum $p_0$ of the particle by operating on $\psi$ with$ \frac{\hbar}{i}d/dx$, more generally with $ \frac{\hbar}{i}\vec\nabla_i$, and thus obtain $p^2_0$, by operating on $\psi$ with $-\hbar^2d^2/dx^2$.  I shall assume this way of obtaining the momentum to be general, and use this assumption in  my derivation of the Heisenberg Uncertainty Relations.

In traditional classical mechanics, where the possibility that the action $S$ may develop caustics is ignored, the continuity equation is also ignored, and thus the space of solutions allowed is larger than allowed here.  Constraints like the continuity equation or (\ref{eq:LaplaceS}) are very strong constraints, and thus quantum mechanics is a strongly constrained classical system.

Notice that we have a derivation of the Copenhagen Interpretation of quantum mechanics.  Recall that Bohr distinguished between ``classical mechanics,''  which described the realm where measurements were recorded, and the world of quantum mechanics.  We have this distinction built in.  The ``classical realm'' is described by those variables for which $\rm R = constant$, and hence trajectories never cross, and the ``quantum realm'' is described by those variables for which $\rm R \not= constant$, and thus the potential $U$ is essential in preventing a breakdown of determinism.  Since for all laboratory experiments, $\rm R$ has compact support, the functions $\psi$ describing these situations are functions in a Hilbert space.

We also have a new means of taking a ``classical limit.''  The usual method of letting $\hbar \rightarrow 0$, is actually invalid, since $\hbar$ is used in the definition of the SI units, and hence letting $\hbar \rightarrow 0$ would physically entail changing the definitions of the units of time, space, and mass.  Papers in high energy physics acknowledge this property of $\hbar$, by setting $\hbar = c = 1$, so taking the limit $\hbar \rightarrow 0$ would be mathematical nonsense in these units, since it would mean letting the integer 1 approach the integer 0: $1\rightarrow 0$ is obviously not allowed.  Instead, the ``classical  limit'' is taken by letting the smoothing potential approach zero, $U \rightarrow 0$, which can be done in a continuous matter by a suitable choice of a continuous sequence of fucntions $R$, allowing $R$ to approach a constant.

The function $\rm R$  can be multiplied by any constant without any physical effect.  As I pointed out above, this is an essential feature required to make equation (\ref{eq:QM}) linear, and thus it is essential to ensure global differentiability.  So for experiments done on objects in the quantum realm, we can set the integral of $\psi^*\psi$ over all the entire quantum configuration space variables equal to one, without loss of generality.  

Erwin Schr\" odinger himself was the first to claim that ${\rm R}^2$ is proportional to the density of universes in the multiverse.  In his English language summary of his new theory, published in 1926, Schr\"odinger wrote that ``$\ldots$ the quantity $\psi\psi^*$ [is] a sort of weight function in the configuration space (\cite{Schrodinger1926PR}, p. 1068).''   Schr\" odinger discovered the Schr\" odinger equation in 1925, and within a year he knew that $\psi$ was a wave, not in ordinary space of three dimensions, but in the $3N$ dimensional configuration space of  $N$ particles.  Operationally, Schr\" odinger treated $\psi^*\psi = {\rm R}^2$ as proportional to a density of the $N$ particles, not in ordinary three-dimensional space, but in the $3N$ dimensional configuration space of these particles.   This is, in all essentials, the meaning of  ${\rm R}^2$ we have been led to via our analysis of Hamilton-Jacobi theory.  Schr\" odinger based his proposal on the experimental fact that transitions from quantum states with a relatively larger value of ${\rm R}^2$ have stronger spectral lines.

Let me now show how this fact, that ${\rm R}^2$ is proportional to the density of universes in the multiverse, implies the Born Interpretation, hence indeed would result in stronger spectral lines.

\section{Deriving the Born Interpretation Using Gibbs Indistinguishability}

There are three distinct types of indistinquishabiity in the H-J equation with Hamiltonian (\ref{eq:generalHamiltonian}) and with smoothing potential.  There are the two familiar forms of indistinguishability, since the smoothing potential and the continuity equation are unchanged if

\begin{eqnarray}
{\rm R}(x_1, x_2, \ldots, x_i, \ldots, x_j, \ldots, x_{3N}, t) = \nonumber\\
 + {\rm R}(x_1, x_2, \ldots, x_j, \ldots, x_i, \ldots, x_{3N}, t) 
\label{eq:BoseEinstein}
\end{eqnarray}
 
 \noindent
 where the $i$th and the $j$th particles of the same mass have been interchanged, and which gives Bose-Einstein Indistinguishability, and 
 
\begin{eqnarray}
{\rm R}(x_1, x_2, \ldots, x_i, \ldots, x_j, \ldots, x_{3N}, t) = \nonumber\\
 - {\rm R}(x_1, x_2, \ldots, x_j, \ldots, x_i, \ldots, x_{3N}, t) 
\label{eq:FermiDirac}
\end{eqnarray}

\noindent
which gives Fermi-Dirac Indistinguishability.  But in both cases, it is assumed that $R$ has compact support, and thus the corresponding $\psi$ is a member of a Hilbert space.

If we consider a particle of mass $M$ for which ${\rm R}$ = constant, then (\ref{eq:BoseEinstein}) still holds, but now the wave function does not yield an attraction between the particles, as it would with Bose-Einstein Indistinguishability.  This is the third type of indistinquishabilty, which, since it appears at the classical level, I shall term {\it Gibbs Indistinguishability}.

It is Gibbs indistinguishability that gives the Born frequencies in repeated measurements.

Consider the experimental apparatus (and experimenter) to be a single classical object of mass $M$, and suppose that before a measurement, the system consists of this classical object, and a separate ``quantum'' system ($\psi$ of compact support) between which there has been no interaction.  Hence before the measurement, the H-J equation with smoothing potential separates.  

For simplicity, assume that the classical apparatus corresponds to $p_0 = 0$, an apparatus at rest.  Then there are an infinity of universes containing such an apparatus, all identical to each other, and hence all collectively subject to Gibbs indistinguishability, since all of these universes have exactly the same mass,  even though they are classical sized objects.

Let us assume, once again for simplicity, that we have the ``quantum'' object to be measured in a superposition of two states.  Let us assume, for example, that we have as our ``quantum'' system to be an electron bound to a proton, and that the two states correspond to $n = 2, \ell = 1, m =  +1$, or $m=-1$.  Let us call these two states respectively``up'' and ``down.''

Arnold Sommerfeld pointed out (\cite{Sommerfeld1930}, p. 6)  in his {\it Wave Mechanics}, the first book length treatise on Schr\"odinger theory, that ''quantum'' mechanics was so named because physicists believed there was an intrinsic granularity to physical reality, expressed in the quantum numbers $n,\ell, m$.  Instead, rather to Sommerfeld's ``surprise,'' these quantum numbers arise from a continuum equation via the boundary conditions on the equation, and thus the quantum numbers are no more an indication of an intrinsic ``quantum'' nature to reality than the appearance of discrete frequencies in a vibrating string which is required to have its two ends fixed. Let us suppose ${\rm R(up)} = \sqrt p$, ${\rm R(down)} = \sqrt q$, and use the fact that we can normalize the superposition to impose $p+q =1$.

Now let the classical system perform a measurement on the ``quantum'' system, and record the measurement, the measurement corresponding to a very weak interaction potential between the two systems.  There will now be a fraction $p$ of the classical universes in which the measurement is up, and a fraction $q$ of the classical universes in which the measurement is down.  But these are the only distinguishing features of these two classes of classical universes.  Because of this Gibbs Indistinquishability, we can equate the relative number of universes $p$ in which the atom is measured to be up to be equal to the probability that we are in one of those universes.  This is the Laplace concept \cite{Jaynes03} of ``probability,'' that if there are $w$ distinct possibilities, each indistinquishable from the others, then the probabilty that we will obtain any one of them is $1/w$.

The probability ${\rm prob}(r|N)$ that an observer in a particular universe will, after $N$ measurements of N different atoms but with all of these $N$ atoms in the state assumed above, with $p + q =1$, see the atom as being in the up state $r$ times, is

\begin{eqnarray}
{\rm prob}(r|N) = \sum_k\,{\rm prob}(r, S_k|N)\nonumber\\
=  \sum_k\,{\rm prob}(r| S_k,\, N)\times {\rm prob}(S_k|N)
\label{eq:prob(r|n)}
\end{eqnarray}

\noindent
where the summation is over all the $2^N$ sequences of outcomes $S_k$, each of which actually occurs in some universe of the multiverse, after $N$ measurements in each of these now $2^N$ distinct universes.  The first term in the second line of (\ref{eq:prob(r|n)}) will equal one if $S_k$ records exactly $r$ measurements of the atom in the $m =+1$  up state, and will be zero otherwise.  Since the $N$ atoms are by assumption independent, the probability of getting any particular sequence $S_k$ depends only on the number of atoms with states measured to be $m = +1$, and on the number with states measured to be $m = -1$.  In particular, since the only sequences that contribute to (\ref{eq:prob(r|n)}) are those with $r$ atoms measured to be in the $m = +1$ up state and those with $N-r$ measured to be in the $m = -1$ down state, we have

 \begin{equation}
{\rm prob}(S_k|N) = p^r q^{N-r}
 \label{eq:prob(S_k|N)}
\end{equation}

However, the order in which the $r$ up states and the $N-r$ down states are obtained is irrelevant, so the number of times (\ref{eq:prob(S_k|N)}) appears in the sum (\ref{eq:prob(r|n)}) will be $C^N_r$, the number of combinations.  Thus the sum (\ref{eq:prob(r|n)}) is

 \begin{equation}
{\rm prob}(r\,|\, N) = \frac{N!}{r!(N-r)!}p^r q^{N-r}
 \label{eq:prob(r|N}
\end{equation}	

The relative number of universes in which we would expect to measure $m = +1$ in $N$ experiments ---that is to say, the expected value of the frequency with which we would measure the atom to be in state $m = +1$ --- is

\begin{equation}
\langle f(m=+1) \rangle =  \sum_{r=0}^N \Big(\frac{r}{N}\Big){\rm prob}(r\,|\, N)\nonumber
\end{equation}

\begin{equation}
=  \sum_{r=1}^N \frac{(N-1)!}{(r-1)!(N-r)!}p^r q^{N-r} = p(p+q)^{N-1} = p
\label{eq:freq}
\end{equation}

\noindent
where the lower limit has been replaced by one, since the value of the $r=0$ term is zero.

The sum in the second line of (\ref{eq:freq}) has been evaluated by differentiating the generating function of the binomial series $\sum_{r=0}^NC^N_r p^rq^{N-r} = (p+q)^N$.  That is, we have $\langle r^m \rangle = (p[d/dp])^m(p+q)^N$, where $q$ is regarded as a constant in the differentiation, setting $p+q = 1$ at the end.  This trick also allows us to show that the variance of the difference between the frequency $f = r/N$ and the probability $p$ vanishes as $N \rightarrow \infty$, since we have

 \begin{equation}
\Big\langle\left( \frac{r}{N} - p\right)^2\Big\rangle = \frac{pq}{N}
 \label{eq:varf}
\end{equation}	
 
\noindent
In fact, all moments of the difference between $f$ and $p$ vanish as $N \rightarrow \infty$, since the generating function gives

 \begin{equation}
\Big\langle\left( \frac{r}{N} - p\right)^{m}\Big\rangle \sim \frac{1}{N} + {\rm higher\, order\,terms\, in }\,\frac{1}{N}
 \label{eq:momentf}
\end{equation}	

So we have

 \begin{equation}
\lim_{N\to\infty}\left( \frac{r}{N}\right) = p
 \label{eq:limtf}
\end{equation}	

\noindent
in the sense that all the moments vanish as $1/N$ as $N \rightarrow \infty$.  This law of large numbers explains why it has been possible to believe, incorrectly, that probabilities are frequencies.   Not so, as Laplace emphasized over two hundred years ago.  It is, instead, that the quantum property of indistinguishability, applied to the observers, forces the measured frequencies to approach the probabilities $p$ and $q$.  The above result generalizes to any number of states, provided only that we limit ourselves to measurements on the part of the H-J equation for which the support is compact, and hence has a wave function that can be regarded as an element in a Hilbert space.

\section{Deriving the Heisenberg Uncertainty Principle}

The Heisenberg Uncertainty Principle does NOT measure any fundamental limitation on our ability to measure a physical quantity, or any limitation on determinism, but rather it is a reflection of the fact that, after we carry out a measurement, we cannot be sure which universe we are in.  There are an infinity of universes, and an infinity of universes which at any instant are identical to each other, by Gibbs indistinquishability.

To see this, I am going to extend an elementary derivation of the Uncertainty Principle, a derivation due to Hermann Weyl.

All derivations of the uncertainty principle concludes the following.  Define the ``expectation value'' of an operator $A$ as

\begin{equation}
\langle A\rangle \equiv \int_{-\infty}^{+\infty}\psi^*A\psi\,dx
\label{eq:quantumexpectationvalue}
\end{equation}	

Notice that this definition of expectation value can be used in the sense defined in the previous section.

Define the ``variance'' $\Delta A$ of an operator $A$ as

\begin{equation}
 \Delta A  \equiv \sqrt{ \langle A^2\rangle - \langle A \rangle^2}
\label{eq:quantumvariance}
\end{equation}	

Then the usual expression of the ``uncertainty principle'' is

\begin{equation}
 \Delta A \Delta B \geq \frac{\hbar}{2}
\label{eq:standarduncertainity}
\end{equation}	

\noindent
where $B$ is some other operator.  Inequality (\ref{eq:standarduncertainity}) will of course not apply to all operators $A$ and $B$, just some pairs of operators. 

A pair of operators to which the uncertainty principle does apply is the operator of the position of a particle, and the operator of the momentum of a particle.  It is this example which is most often given when the ``limitation of measurement'' claim is made, so I shall focus on this case.  The Weyl proof, which is described in (\cite{Landau77}, p. 48) deals with this case, and restricts itself to just one spatial dimension, chosen to be the $x$ location of the particle.  For simplicity, Weyl also assumes that $\langle x \rangle = 0$, and $\langle p \rangle = 0$, where $p$ is the momentum of the particle in the $x$ direction, so we only have to show that if

\begin{equation}
\langle x^2\rangle \equiv \int_{-\infty}^{+\infty}\psi^*x^2\psi\,dx
\label{eq:quantumexpectationvaluexsquarex}
\end{equation}	

\noindent
and

\begin{equation}
\langle p^2\rangle \equiv \int_{-\infty}^{+\infty}\psi^*p^2\psi\,dx
\label{eq:quantumexpectationvaluexsquarep}
\end{equation}	

\noindent
then

\begin{equation}
( \Delta x)^2( \Delta p)^2 \geq \left(\frac{\hbar}{2}\right)^2
\label{eq:standarduncertainityxp}
\end{equation}	

Weyl's proof begins with the inequality

\begin{equation}
\int_{-\infty}^{+\infty}\left|\alpha x\psi + \frac{d\psi}{dx}\right|^2\,dx \geq 0
\label{eq:Weylinequality}
\end{equation}	

\noindent
where, as usual, $|Q|^2 \equiv Q^*Q$.  The number $\alpha$ is assumed to be any real number.

When expanded out algebraically, the integral becomes the sum of three integrals, an integral in $x^2|\psi|^2$, an integral involving cross terms in $x\psi d\psi/dx$,and an integral in $|d\psi/dx|^2$.  The cross term integral can be evaluated as 

\begin{equation}
\int_{-\infty}^{+\infty}\left(x\frac{d\psi^*}{dx}\psi + x\psi^*\frac{d\psi}{dx}\right)\,dx = \int_{-\infty}^{+\infty}x\frac{d|\psi |^2}{dx}\,dx
\label{eq:Weylintegralmixed}
\end{equation}	

We integrate by parts of last integral in (\ref{eq:Weylintegralmixed}).   We note that $|\psi|^2$ vanishes at infinity (otherwise the integral of $|\psi|^2$ over all space would not be finite,and hence would not be a ``quantum'' state, since it would not be an element of a Hilbert space).  Combining these calculations give

\begin{equation}
\int_{-\infty}^{+\infty}\left(x\frac{\psi^*}{dx}\psi + x\psi^*\frac{d\psi}{dx}\right)\,dx = - \int_{-\infty}^{+\infty}|\psi|^2\, dx = -1
\label{eq:Weylintegralmixed2}
\end{equation}	

The integral with integrand $|d\psi/dx|^2$ is

\begin{equation}
\int_{-\infty}^{+\infty}\frac{d\psi^*}{dx}\frac{d\psi}{dx}\,dx = - \int_{-\infty}^{+\infty}\psi^*\frac{d^2\psi}{dx^2}\, dx
\label{eq:Weylintegraltwoderivatives}
\end{equation}	

\noindent
where once again integration by parts has been used.  

Now the operators for $x^2$ and $p^2$ are $x^2$ and $(-\hbar^2)d^2/dx^2$ respectively, the latter being justified in Section I.  Thus we have

\begin{equation}
( \Delta x)^2 = \int_{-\infty}^{+\infty}x^2\psi^*\psi\, dx \equiv \int_{-\infty}^{+\infty}x^2|\psi|^2\, dx
\label{eq:varianceforx}
\end{equation}	

\noindent
and

\begin{equation}
( \Delta p)^2 = \int_{-\infty}^{+\infty}\psi^*p^2\psi\, dx = -\hbar^2 \int_{-\infty}^{+\infty}\psi^*\frac{d^2\psi}{dx^2}\, dx
\label{eq:varianceforp}
\end{equation}	

The last integral in (\ref{eq:varianceforp}) is just the last integral in (\ref{eq:Weylintegraltwoderivatives}), multiplied by $\hbar^2$.  Thus we can write the ```obvious inequality'' (\ref{eq:Weylinequality}) as

\begin{equation}
\alpha^2(\Delta x)^2 - \alpha + \frac{1}{\hbar^2}(\Delta p)^2 \geq 0
\label{eq:Weylinequality2}
\end{equation}	

Think of the left hand side of this inequality as a function of $\alpha$:

\begin{equation}
 g(\alpha) \equiv \alpha^2(\Delta x)^2 - \alpha + \frac{1}{\hbar^2}(\Delta p)^2
\label{eq:Weylinequality3}
\end{equation}	

We can find the value of $\alpha$ that minimizes the function $g(\alpha)$ by setting $dg/d\alpha = 0$ and $d^2g/d\alpha^2  >0$, which gives

\begin{equation}
\alpha =\frac{1}{2(\Delta x)^2}
\label{eq:Weylinequality4}
\end{equation}	

Inserting (\ref{eq:Weylinequality4}) into (\ref{eq:Weylinequality2}), and a little algebra gives

\begin{equation}
( \Delta x)^2( \Delta p)^2 \geq \left(\frac{\hbar}{2}\right)^2
\label{eq:standarduncertainityxp2}
\end{equation}	

\noindent
which is just the uncertainty relation (\ref{eq:standarduncertainityxp}).  

Notice that the Weyl derivation of the uncertainty principle makes no reference to any measurement.  No rigorous derivation does.  Therefore, the uncertainty principle cannot be a limitation on measurement.

The cause of the uncertainty principle is not a limitation on measurement, but rather an interaction of the other universes of the multiverse with our universe.  To see this, solve the Schr\" odinger equation (\ref{eq:QM}) for the second derivative of $\psi$, and substitute this into the last integral of (\ref{eq:Weylintegraltwoderivatives}).  Then calculate $\partial\psi/\partial t$ by using the expression (\ref{eq:Pilotwave}) for $\psi$.  The result is a sum of three integrals, one of which must vanish, since otherwise it would give an imaginary contribution to (\ref{eq:Weylintegraltwoderivatives}), whereas we know this expression is purely real.  Then equation (\ref{eq:Weylintegraltwoderivatives}) is also equal to

\begin{equation}
 \int_{-\infty}^{+\infty}|\psi|^2\left(\frac{1}{\hbar^2} \left(  \frac{\partial { S}}{\partial x}\right)^2 + \frac{\nabla^2{\rm R}}{{\rm R}}\right)\,dx
\label{eq:twoderivativesHJ}
\end{equation}	

Thus we can write the variances of both position and momentum in the same form for comparison:

\begin{equation}
( \Delta p)^2 =  \int_{-\infty}^{+\infty}|\psi|^2\left(\left(  \frac{\partial { S}}{\partial x}\right)^2 + \hbar^2\frac{\nabla^2{\rm R}}{{\rm R}}\right)\,dx
\label{eq:varianceforpHJ}
\end{equation}	

\noindent
and

\begin{equation}
( \Delta x)^2 =  \int_{-\infty}^{+\infty}|\psi|^2x^2\, dx
\label{eq:varianceforx2}
\end{equation}

This allows us to see the origin of the uncertainty principle.  In Hamilon-Jacobi theory without the smoothing potential term, the momentum of a particle is just $p = \partial {S}/\partial x$.  If there were only one universe this would mean that the density of universes would be zero except for a single point.  This implies that $|\psi|^2$ is a delta function.  Recall that a delta function $\delta(x - x_0)$ is defined by

\begin{equation}
 \int_{-\infty}^{+\infty}\delta(x-x_0)f(x)\, dx = f(x_0)
\label{eq:deltafunction}
\end{equation}	

\noindent
where $x_0$ is a constant, and $f(x)$ is any function.  In the above derivation of the uncertainty principle, we set the expectation values of both the position and the momentum to be zero to simplify the mathematics.  With a single universe, this would mean setting $|\psi|^2 = \delta(x - x_0) =\delta(x)$, and $p= \partial {S}/\partial x =0$ at $x=0$.  Equations (\ref {eq:varianceforpHJ}) and (\ref{eq:varianceforx2}) then give $(\Delta x)^2 = (\Delta p)^2 = 0$, which is to say, we would have both variances equal to zero simultaneously, if the smoothing potential vanishes. 

But mathematical consistency requires the smoothing potential be non-zero.  This means that we can no longer set $|\psi|^2 = \delta(x-x_0)$, because if we did, the smoothing potential, which is proportional to the second derivative of $|\psi| = {\rm R}$, would make the integral in (\ref{eq:varianceforpHJ}) infinite.  Now since $|\psi|^2$ is proportional to the density of universes in the multiverse, this means that the ultimate reason for the uncertainty principle is not a limitation on our ability to measure position and momentum simultaneously, but rather that any attempt to measure these quantities with absolute precision in one universe would increase to infinity the interference with our measurements from the other universes.

\section{Conclusion}

Schr\"odinger's equation (\ref{eq:QM}) is equivalent to another action principle, the Feynman path integral \cite{FeynmanHibbs1965}:

\begin{equation}
\psi(x_f, t_f) = \int^{+\infty}_{-\infty}K(x_f,t_f, x_i,t_i)\psi(x_i, t_i)d\,x_i
\label{eq:wavefunctionpropaged}
\end{equation}	

\noindent
where the Feynman propagator $K(x_f,t_f, x_i,t_i)$ is defined in terms of a path integral:

\begin{equation}
K(x_f,t_f, x_i,t_i) = \int^{t_f}_{t_i}{\rm exp}\left[\frac{i}{\hbar}S(x)\right]Dx
\label{eq:pathintegral}
\end{equation}

The path integral uses the intrinsic linearity of (\ref{eq:QM}) to superpose, with a weight function ${\rm exp}\left[\frac{i}{\hbar}S(x)\right]$ along each individual path,  all possible paths from an initial multiverse to a final multiverse.  As we saw in Section I, all paths are included even in the classical H-J equation, but only the smoothing potential (\ref{eq:quanpot}) forces the paths to be consistent with $C^2$ assumption of classical mechanics.  In (\ref{eq:pathintegral}), no smoothing potential is necessary; $S(x)$ is the action with $U=0$.  The sum over all histories --- over all possible universes of the multiverse --- is mathematically equivalent to adding a smoothing potential.

I have shown that the Copenhagen interpretation is automatically included in the derivation of quantum mechanics from H-J theory:  the ``classical'' regime of the Copenhagen Interpretation are those for which the smoothing potential vanishes, and thus the wave functions for such states are not elements of a Hilbert space. 

Needless to say, the Many-Worlds Interpretation (\cite{Everett1957}, \cite{DeWittGraham1973}) is also included. However, I have shown that the Many-Worlds are not a purely quantum phenomenon, but rather merely an automatic consequence of H-J theory.  Further, the standard Many-Worlds theory starts with the completely unnecessary assumption that the allowed function space for the wave function is a Hilbert space, thus making it hard to see why $\psi\psi^*$ is proportional to the density of universes in the multiverse.  Using the approach of this paper, this proportionality falls out of the mathematics.  And as a bonus, we see why we cannot do better than ``proportional to.''  I have shown the very existence of an attractive potential forces us to accept the existence of the other universes of the multiverse, since the attractive potential shows that these other worlds can interact with our own world.  In Section II, the Born frequencies were shown to arise as a consequence of the real existence of the other universes; the experimental confirmation of the Born frequencies should be regarded as experimental evidence for their existence.

In spite of this interaction between the universes, the existence of the other universes has been claimed to be a violation of Ockham's Razor: {\it pluralitas non est ponenda sine necessitate} (``plurality should not be posited without necessity'').  Remarkably, we have the written opinion of none other than William of Ockham himself that this is not so, and would not be so even if the other universes did not interact with our own!

In the Medieval world-view, the phrase ``plurality of worlds'' denoted what we would now call a ''plurality of universes'' (\cite{Duhem1985}, p. 452).  A ``universe'' was a an Earth-centered whole, surrounded by a sphere of fixed stars, and outside this ultimate sphere there was nothing: no matter and no empty space.  William of Ockham argued that these other universes could exert no effect  whatsoever on our own universe (\cite{Duhem1985}, p. 462--464).  Ockham explicitly claimed that the existence of these universes, which he believed were totally undetectable by us, could nevertheless not be ruled out, in spite of what Ockham's Razor would appear to dictate.

Ockham said that his Razor could not eliminate the existence of undetectable universes if their existence was required by a theory we knew to be true on other grounds.  For Ockham, the existence of an omnipotent God was a fact, and an omnipotent God could make undetectable universes if He wanted to do so.  Arthur Lovejoy (\cite{Lovejoy1964}, p. 52, 74) has pointed out that the most basic postulates of Medieval theory implied that God would necessarily create all possible such universes, so an infinity of them would exist, undetectable by us.

But of course we are not in the position of Ockham, since, as shown in the previous sections, the other universes of the multiverse do in fact interact with ours.  In fact, as shown in Section II, measurement of the Born frequencies is actually a measurement proving the existence of the other classical universes.  As shown in Section III, confirmation of the Heisenberg Uncertainty Principle is also a confirmation of the interaction of the other universes interacting with our own.

If we had not added the smoothing potential, we would have begun with the assumption that physical quantities were everywhere $C^2$, and from this concluded that physical quantities were not everywhere $C^2$ --- which is to say, we would have a logical contradiction without the smoothing potential. A logical contradiction is the worse possible violation of Ockham's Razor, since anything at all can be deduced from a contradiction.

The derivation of Schr\"odinger's equation from the H-J equation gives the natural function space for quantum mechanics.  It is not necessary to postulate that $\psi$ is an element of a Hilbert space.  In fact, the natural function space is indeed much larger, as pointed out in Section I.  We need no longer have to extend the function space to include the extremely useful plane wave states.  They are already in the natural function space.  Feynman pointed out that the propagator defined by his path integral was a solution to the Schr\"odinger equation, but he did not consider the propagator to be an acceptable wave function, since it cannot be an element of a Hilbert space.  This is now changed, and I have argued in \cite{Tipler2005} that having such a wave function as the wave function of the universe would automatically solve the Flatness Problem in cosmology, using the quantum kinematics of wave packet spreading.

We can, if we wish, widen the function space even more, and include the non-interacting worlds of Ockham.  We can add as many sets of $3N$ particles as we wish, and as long as the two potentials have no interaction between the sets, the solutions for this more general H-J equation will separate, and we will have a multiverse in which we indeed have non-interacting universes, in general with different physical laws (different potentials in such universes).  But I am an old fashioned physicist, and, unlike William of Ockham, I deny the existence of any entity that cannot, even in principle, interact with the world of common experience.

But the universes of the H-J multiverse do interact with each other.  Measurement of the Born frequencies and confirming the Uncertainty Relations are observations of this interaction.  We do not usually think of these measurements as showing interaction between the universes, but in Medieval physics, neither was the rising and setting of the Sun considered evidence for the Earth's rotation.  Newtonian physics says that, nevertheless, such observations are evidence for the other universes, and the Earth's rotation, respectively.  We also see that the Copenhagen and the Many-Worlds Interpretations are not competing interpretations, but complementary interpretations, emphasizing different aspects of physical reality, which is ultimately classical mechanics made globally $C^2$ and hence globally deterministic.

\end{document}